\documentclass[11pt]{article}
\usepackage{amsmath,amssymb,color,graphics,epsfig,cite}
\usepackage{hyperref}
\usepackage{ulem}

\usepackage{amsfonts}

\newcommand{\be}{\begin{equation}}
\newcommand{\ee}{\end{equation}}
\newcommand{\bea}{\setlength\arraycolsep{2pt} \begin{eqnarray}}
\newcommand{\eea}{\end{eqnarray}}
\newcommand{\nn}{\nonumber}

\def\ft#1#2{{\textstyle{\frac{\scriptstyle #1}{\scriptstyle #2} } }}
\def\fft#1#2{{\frac{#1}{#2}}}

\def\0{{\sst{(0)}}}
\def\1{{\sst{(1)}}}
\def\2{{\sst{(2)}}}
\def\3{{\sst{(3)}}}
\def\4{{\sst{(4)}}}
\def\5{{\sst{(5)}}}
\def\6{{\sst{(6)}}}
\def\7{{\sst{(7)}}}
\def\8{{\sst{(8)}}}
\def\sst#1{{\scriptscriptstyle #1}}

\begin{document}

 
\begin{center}
		{\Large {\bf Hairy Black Holes with Arbitrary Small Areas}}

		\vspace{40pt}
  
{\large{Xiao-Ping Rao$^{1}$, Hyat Huang$^{1,2}$, and Jinbo Yang$^3$}}

\vspace{10pt}

{\it $^1$College of Physics and Communication Electronics, Jiangxi Normal University, Nanchang 330022, China\\
$^2$  Institute of Physics, University of Oldenburg, Postfach 2503, D-26111 Oldenburg,
Germany\\
$^3$ Department of Astronomy, School of Physics and Materials Science,
		Guangzhou University, Guangzhou 510006,China
}

		\vspace{40pt}
		
		\underline{ABSTRACT}
	\end{center}
	
We obtained new hairy black hole solutions in Einstein-scalar theory, including asymptotic flat, de Sitter and anti-de Sitter black holes. The theory is inspired by Ref.~\cite{Huang:2020qmn}, where traversable wormhole solutions from an Einstein-phantom scalar theory are constructed. In this work, we found new black hole solutions in an  Einstein-normal scalar theory. Comparing with Schwarzschild metric, the hairy black holes have two interesting properties: i) the areas of the black holes are always smaller than the same mass Schwarzschild black holes; ii) A naked singularity with positive mass arises when the black hole mass decreases. The energy conditions for the black holes and naked singularities are checked. We found that, as hairy black holes, the null energy condition(NEC) and the strong energy condition(SEC) are hold, while the weak energy condition(WEC) is violated in the vicinity of black hole horizon. The naked singularity respects to all three energy conditions. We also investigate the quasinormal modes(QNMs) of the hairy black holes by a test scalar field. The results indicate that one can distinguish hairy black holes with the same mass Schwarzschilid black hole by their QNM spectra.

\vfill {\footnotesize 
~\\
xiaopingrao@jxnu.edu.cn\\
hyat@mail.bnu.edu.cn\\
yangjinbo@gzhu.edu.cn

}\ \ \ \

	\thispagestyle{empty}
	
	\pagebreak
\addtocontents{toc}{\protect\setcounter{tocdepth}{2}}
	


\section{Introduction}

The study of black holes is quite prominent nowadays because of the shadow images cast by the M87* black hole\cite{EventHorizonTelescope:2019dse} and the Sagittarius A* black hole \cite{EventHorizonTelescope:2022wkp}. On the other hand, the detection of gravitational waves (GWs) originated from the merger of binary black holes in 2015 laid the conceptual cornerstone for gravitational detections \cite{LIGOScientific:2016aoc}. These great progresses in experiments provide fruitful data to the theoretical studies of black holes and hence have launched an extraordinary new era in black hole physics. Therefore, more and more gravitational experiments are being planned or are already underway, such as the LISA in Europe/USA, KAGRA in Japan, GEO600 in Germany, Taiji and Tianqin in China \cite{KAGRA:2013rdx,Somiya:2011np,Berti:2005ys,Caprini:2019egz,Luck:1997hv,Hu:2017mde,Ruan:2020smc,TianQin:2015yph,Huang:2020rjf}.

In theory, the no-hair conjecture plays a pivotal role in black hole physics. The uniqueness of Schwarzschild black hole and Kerr black hole has been demonstrated (see Ref.~\cite{Israel:1967wq,Robinson:1975bv,Bekenstein:1996pn,Cardoso:2016ryw,Yang:2023nnk}), which state that one can describe a stationary black hole solely in terms of its mass $M$, and angular momentum $J$. The idea of testing the no-hair conjecture with black hole shadow images and GWs has caused widespread concern \cite{Cunha:2015yba,Khodadi:2020jij,Afrin:2021imp,Vagnozzi:2022moj,Khodadi:2021gbc,Li:2023zbm,Gossan:2011ha,Sampson:2014qqa,Isi:2019aib,Bhagwat:2019dtm}. However, we can't rule out the existence of hairy black holes yet. In fact, many black holes with scalar hair \cite{Bekenstein:1975ts,Bizon:1994dh,Nucamendi:1995ex,Volkov:1998cc,Radu:2011uj,Anabalon:2013qua,Feng:2013tza,Herdeiro:2014goa,Sotiriou:2014pfa,Fan:2015oca,Herdeiro:2015gia}, Yang-Mills hair\cite{Luckock:1986tr,Kleihaus:1997rb,Kleihaus:1998sm,Hartmann:2001ic,Kleihaus:2002ee,Ibadov:2005rb}, and so on\cite{Brito:2013xaa,Herdeiro:2016tmi,Fan:2016jnz,Shnir:2020hau} have been constructed. These hairy black holes are so attractive that there are numerous theoretical studies on their properties \cite{Tamaki:2001xf,Lu:2014maa,Fan:2016yqv,Guo:2020caw} and experimental constraints \cite{Berti:2015itd,Cunha:2015yba,Barack:2018yly,Gogoi:2024vcx}. 

Black hole with scalar hair, namely imposing a scalar field into the Einstein gravity, is the simplest example of hairy black holes. 
Although the only scalar degree of freedom in the Standard Model (SM) is the Higgs field, scalar fields are utilized in numerous aspects of fundamental physics, like the dark matter (DM) candidates including axion-like particles (ALP)\cite{Contino:2013kra,Preskill:1982cy,Feng:2010gw,Caldwell:2016dcw,Ballesteros:2016euj,Hui:2016ltb}, the nature of dynamical dark energy\cite{Copeland:2006wr,Li:2011sd}, and inflatons in many usually considered inflation models\cite{Kofman:1994rk,Martin:2013tda}. As in string theory, scalar fields appear in several contexts:
the so-called dilaton generally appears as a kind of variation mode on a string; the moduli massless scalars or axions from higher-form fields emerge when compactifying six extra dimensions\cite{Svrcek:2006yi,Arvanitaki:2009fg,Palti:2019pca}.
Despite Bekenstein's investigation from 1971 on the General Relativity minimally coupled to a scalar field, which resulted in a no-scalar hair theorem (originating in \cite{Bekenstein:1972ny,Bekenstein:1971hc,Bekenstein:1972ky,Bekenstein:1995un} and reviewed in \cite{Herdeiro:2015waa}). It is possible to circumvent some of the assumptions he made to obtain hairy black holes. Such as by introducing higher-order derivative gravities like the Gauss-Bonnet term, non-minimally coupled to the scalar field. This may give rise to an intriguing phenomenon known as ``black hole scalarization”, wherein non-hairy black holes can evolve and spontaneously develop scalar hair themselves\cite{Blazquez-Salcedo:2018jnn,Berti:2020kgk,Kleihaus:2015iea,Lai:2022spn,Lai:2022ppn,Fan:2023jhi}. 

Another path to bypass the no-scalar hair theorem is by violating the energy conditions, especially the weak energy condition. In around 1970s, R. Penrose and S. Hawking came up with a series of works on singularity theorems, which require adherence to the energy conditions \cite{Penrose:1964wq,Penrose:1969pc,Hawking:1970zqf}. However, in both theoretical and experimental research, there remains a strong desire towards matters that violate energy conditions. A well-known example in quantum field theory is the Casimir effect, where the null energy condition could be violated\cite{Graham:2006mx,Alexandre:2023iig}. Based on cosmological observations, our Universe is undergoing accelerated expansion\cite{SupernovaSearchTeam:1998fmf,SupernovaCosmologyProject:1998vns,Perlmutter:2003kf}. The cosmological constant can be treated as a dark energy model that supports this accelerated expansion and violates weak energy condition. Furthermore, the existence of exotic objects such as traversable wormholes also necessitates the violation of energy conditions\cite{Morris:1988cz,Morris:1988tu,Huang:2019arj,Yang:2021diz}.

In this work, we examine the Einstein-scalar theory and construct hairy black hole solutions. The theory contains a scalar field with a potential inspired by Ref.~\cite{Huang:2020qmn}. We obtain hairy black holes whose asymptotic behavior can be (A)dS or Minkowski spacetime depending on the parameters. (A)dS black holes are attractive in holographic theories, but in astrophysics, asymptotically flat black holes are more realistic. We mainly discuss the asymptotic black hole solutions in this work. We find that the areas of hairy black holes are always smaller than those of Schwarzschild black holes with the same mass. As the scalar parameter $q$ approaches a critical value, the black hole area can become infinitesimally small, which may have significant observational implications. Hence, we examine their QNM's spectra compared to those of Schwarzschild black holes with the same mass by evolving test scalar fields. The results show that the hairy black holes possess higher
oscillation frequencies and shorter decay time. On the other hand, the black hole solutions can become naked singularities as the scalar parameter $q$ varies. We find that such naked singularities can satisfy null, strong, and weak energy conditions. However, the black hole solutions always violate the weak energy condition and thus do not violate the no-scalar hair theorem in Ref.~\cite{Bekenstein:1995un}. Our study provides an analytical model for hairy black hole studies.

Our works are organized as follows. In section 2, we present the theory and the analytical black hole solution. Then we discuss the properties of the black holes in section 3 and examine the energy conditions in section 4. The QNMs of the test field are investigated in section 5. Finally, we conclude our work in section 6 and the appendix in the transformation to obtain the theory.

\section{The theory and the solution}\label{se1}

The Lagrangian for Einstein-scalar theories which is to be considered in our work is given by
\be\label{lag}
{\cal L}=\sqrt{-g}(R-\ft 1 2 (\partial \phi)^2-V(\phi) ),
\ee
where $R$ is the scalar curvature, and $(\partial \phi)^2$ is $g^{\mu\nu}\partial_\mu\phi\, \partial_\nu\phi$ for short. While the scalar potential is
\be
V(\phi)=\alpha \,U(\phi)+\beta\, U(-\phi)
\ee
with
\be
 U(\phi)=e^{-\sqrt{\gamma ^2-1}\, \phi }
\bigg((2-3\gamma^2)\,\text{cosh}(\gamma\phi)-3\gamma\sqrt{\gamma^2-1}\,\text{sinh}(\gamma\phi)-2\bigg). \label{NozawaPoten}
\ee
The coefficients $\gamma$, $\beta$ and $\alpha$ in the scalar potential are three theoretical parameters with $\gamma>1$. Note that the potential \eqref{NozawaPoten} has the similar form to the Eq.(63) of Ref.~\cite{Nozawa:2020gzz} in $\phi\to i\phi$.

The equations of motion (E.O.Ms) associated with the variation of $\phi$ and $g^{\mu\nu}$  are respectively given by
\bea\label{eom}
&&\Box\phi =\fft{\partial V}{\partial \phi} \,,\cr
&&E_{\mu\nu} \equiv R_{\mu\nu}-\ft 12 R g_{\mu\nu}-T_{\mu\nu}^{\phi}=0\,,
\eea
where
\be\label{EMT}
T_{\mu\nu}^{\phi} =\bigg(\ft 12\partial_\mu\phi \partial_\nu \phi-\ft 14 g_{\mu\nu} (\partial \phi)^2\bigg)-\ft 1 2g_{\mu\nu}V.
\ee
We consider the spherically symmetric and static ansatz
\be
ds^2=-h dt^2+f^{-1}dr^2+R^2d\Omega_2^2,\qquad \phi=\phi(r),
\ee
where $d\Omega^2_2=d\theta^2+\sin^2\theta d\varphi^2$ the $2$-sphere and $(h,f,R)$ \footnote{We should not mix the metric function and the Ricci scalar, even though we use the same symbol $R$ to denote them.} are the metric functions of $r$. Substitutes this ansatz into \eqref{eom} gives four equations, which are given by
\bea\label{eomm} 
&&-4+4fR'^2+R^2(2V+f\phi'^2)+4R(2fR''+f'R')=0,\nn\\
&&-4h+4hfR'^2+2hR^2V-hfR^2\phi'^2+4fRR'h'=0,\nn\\
&& R(2hfh''+h^2(f\phi'^2+2V))+R(hh'f'-fh'^2)+2R'(fhh'+h^2f')+4h'fR''=0,\nn\\
&&R\phi'(hf)'+2hf(2R'\phi'+R\phi'')-2hR\fft{\partial V}{\partial \phi}=0, 
\eea
where a prime denotes the derivative with respect to $r$. It is easy to find that the scalar equations of motion are automatically satisfied provided that the above equations are satisfied, and hence only three independent equations left here. It's straightforward to check the following solution satisfies all the E.O.Ms \eqref{eomm}
\bea\label{solution}
ds^2&=&-h(r) dt^2+ f(r)^{-1}dr^2+R(r)^{2}d\Omega_{2}^2, \qquad R(r)^{2}=e^{-\sqrt{\gamma^2-1}\phi}r^2, \qquad f=h\fft{r^2+q^2\gamma^2}{r^2},\nn\\
h&=& \fft{\alpha}{2} e^{-\sqrt{\gamma ^2-1} \phi }\gamma^2r^2+\fft{\beta}{2}e^{\sqrt{\gamma ^2-1} \phi }\gamma^2\big(r^2+q^2\gamma^2-4q\sqrt{r^2+q^2\gamma^2}\sqrt{\gamma^2-1}+q^2(7\gamma^2-8)\big)\nn\\
&&+e^{\sqrt{\gamma 
 ^2-1} \phi },\nn\\
\phi &=&\fft{2}{\gamma}\text{arctanh}(\fft{q\gamma}{\sqrt{r^2+q^2\gamma^2}}),
\eea  
and hence a solution of the Einstein-normal scalar theory\footnote{This solution can be obtained by a disformal transformation of a special Einstein-phantom scalar theory, see appendix A.}. Analysis of the metric function yields the solution represents single-horizon black holes or time-like naked singularities. We expand the metric functions at $r\to\infty$,
\bea\label{exi} 
h&=&-\fft{\Lambda}{3}r^2+\fft{2}{3}\Lambda \sqrt{\gamma^2-1}\, q r+\big(1-\fft{2}{3}\Lambda (\gamma^2-1)\,q^2\big)\nn\\ && \quad+\fft{q\sqrt{\gamma^2-1}\bigg(18-q^2(3\gamma^2-4)(12\alpha\gamma^2+7\Lambda)\bigg)}{9r}+...\,,\nn\\
f&=&-\fft{\Lambda}{3}r^2+\fft{2}{3}\Lambda \sqrt{\gamma^2-1}\,
q r+\big(1-\Lambda (\gamma^2-\fft{2}{3} )\,q^2\big)+...\,,\nn\\
R^2&=&r^2-2qr\sqrt{\gamma^2-1}+2q^2(\gamma^2-1)+...\,,
\eea
where $\Lambda$ is the effective cosmological constant, defined as
\be
 \Lambda \equiv -\fft{3}{2}(\alpha+\beta)\gamma^2 \nn\,.
\ee
It shows that the solution is asymptotic (A)dS or Minkowski spacetime depending on the value of $\Lambda$. AdS black holes can be served as concrete models of AdS/CFT and dS black holes are useful in the context of cosmology.
However, as we mentioned in the introduction, our interest of this work is mainly in asymptotic flat black hole solutions.

\section{Asymptotic flat black holes}

\begin{figure}[t]
\centering
\includegraphics[width=0.3\textwidth]{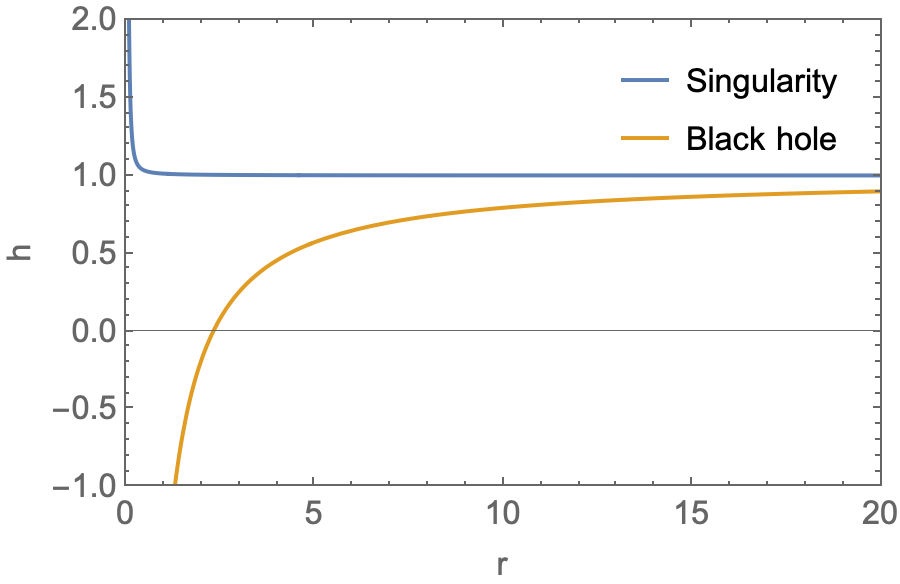} \qquad
\includegraphics[width=0.3\textwidth]{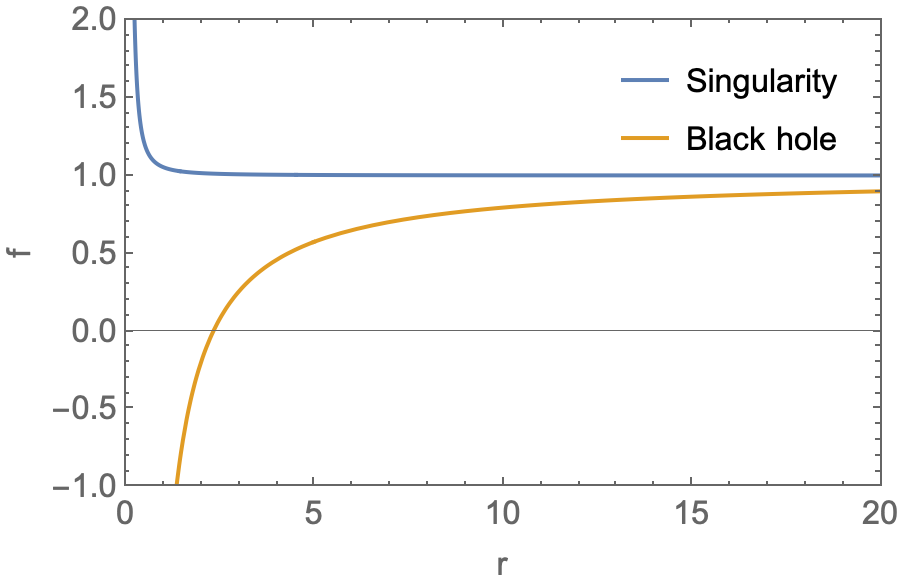} 
\includegraphics[width=0.3\textwidth]{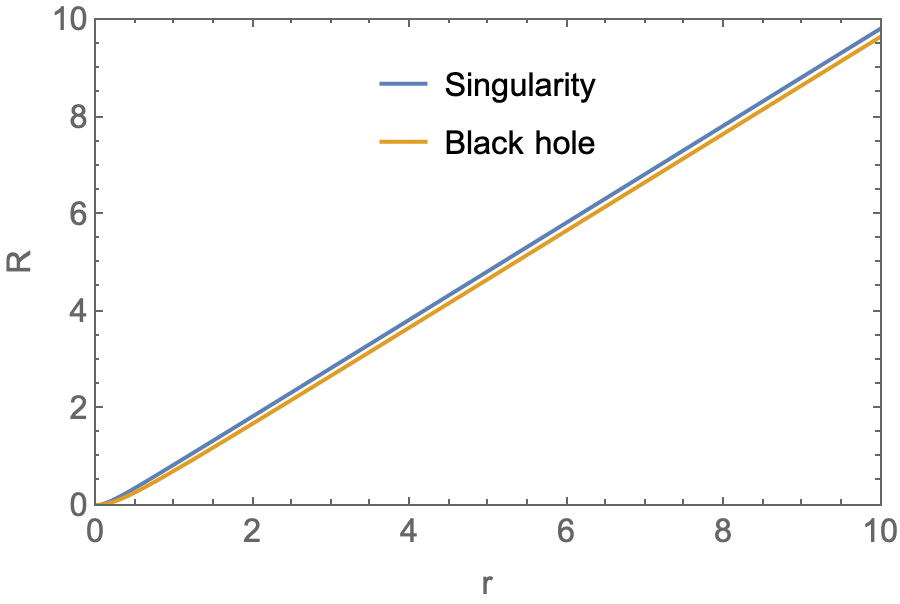} \qquad
\caption{ \it Examples of the metric functions $(h, f, R)$.}
\label{metric}
\end{figure}

The asymptotic flat solution arises in $\alpha+\beta$=0. Without loss of the generity, we set $\beta=-\alpha$, throughout the rest paper. According to \eqref{exi}, the mass of the solution can be read as 
\be\label{mass}
M=\fft{q\sqrt{\gamma^2-1}\bigg(2q^2\alpha\gamma^2(3\gamma^2-4)-3\bigg)}{3}.
\ee
The positiveness of the mass requires
\begin{itemize}
    \item $\alpha q^2 >\fft{3}{2\gamma^2(3\gamma^2-4)}$, if $\gamma>\fft{2}{\sqrt{3}}$;
    \item $\alpha q^2 < \fft{3}{2\gamma^2(3\gamma^2-4)}$, if $1<\gamma<\fft{2}{\sqrt{3}}$.
\end{itemize}
At $r\to0$, we have 
\be
h(0)\sim \infty\bigg(1-2\alpha q^2 \gamma^2(2\gamma^2-2-\gamma\sqrt{\gamma^2-1})\bigg)+\dots .
\ee
With respect to the positive mass condition, and make sure $h(0)$ is negative leads to black hole solutions, namely
\begin{itemize}
    \item $\alpha q^2 >\ft{-1}{2(2-2\gamma^2+\gamma\sqrt{\gamma^2-1})}$, if $\gamma>\ft{2}{\sqrt{3}}$;
    \item $\alpha q^2 < \ft{-1}{2(2-2\gamma^2+\gamma\sqrt{\gamma^2-1})}$, if $1<\gamma<\ft{2}{\sqrt{3}}$.
\end{itemize}
Otherwise, it is a time-like naked singularity. For example, it represents a black hole with a horizon at $r_h=2.315$ and unit mass by setting $\gamma=2, q=0.2, \alpha=4.555$. We show the metric function in Fig.\ref{metric}. Another example is a time-like naked singularity, where $\gamma=2, q=0.1, \alpha=4.555$, as also shown in Fig.~\ref{metric}.

\begin{figure}[h]
\centering
\includegraphics[width=0.4\textwidth]{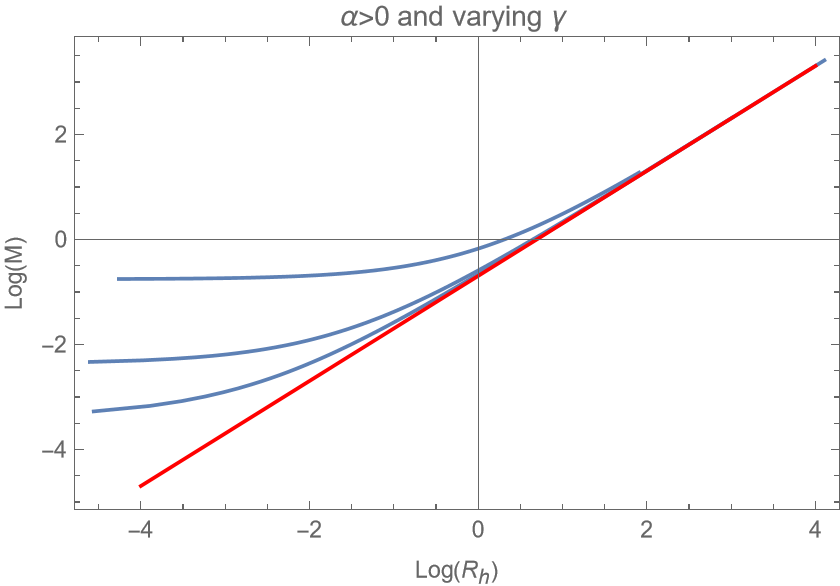} 
\includegraphics[width=0.4\textwidth]{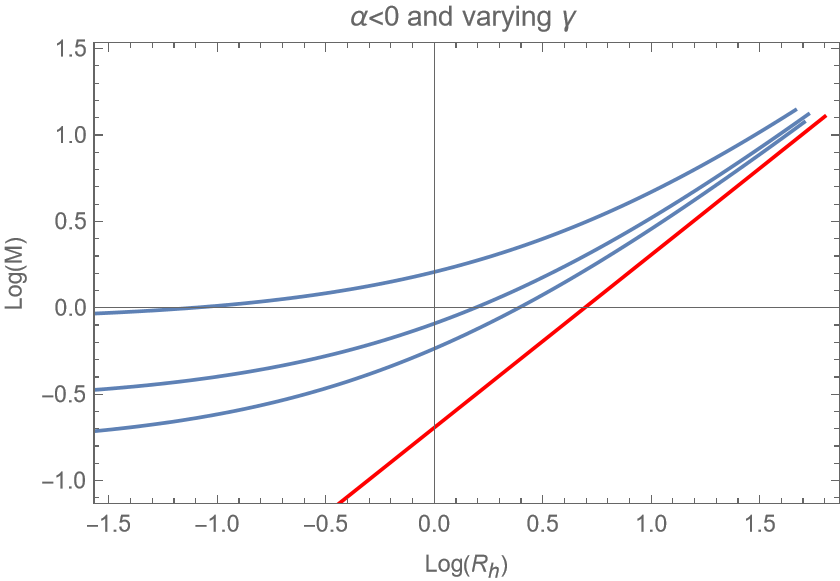} 
\includegraphics[width=0.4\textwidth]{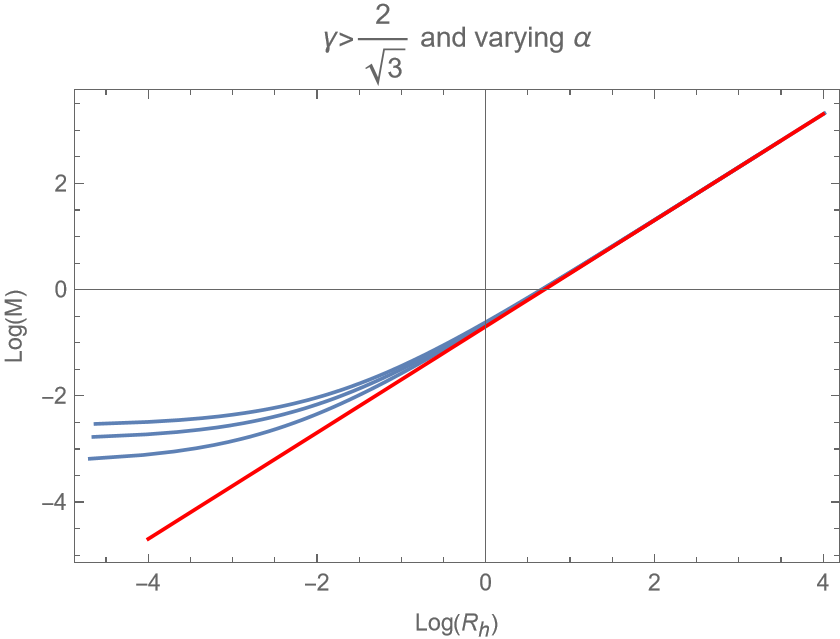} 
\includegraphics[width=0.4\textwidth]{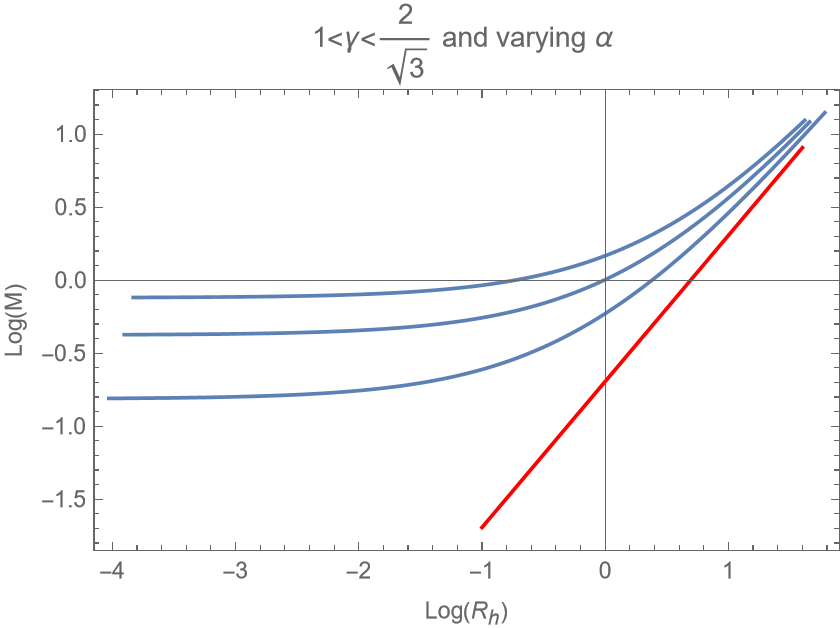} 
\caption{ \it The relations between black hole mass and radius. The red lines denote standard Schwarzschild black holes and the blue lines denote hairy black holes.}
\label{rm}
\end{figure}

\begin{figure}[h]
\centering
\includegraphics[width=0.4\textwidth]{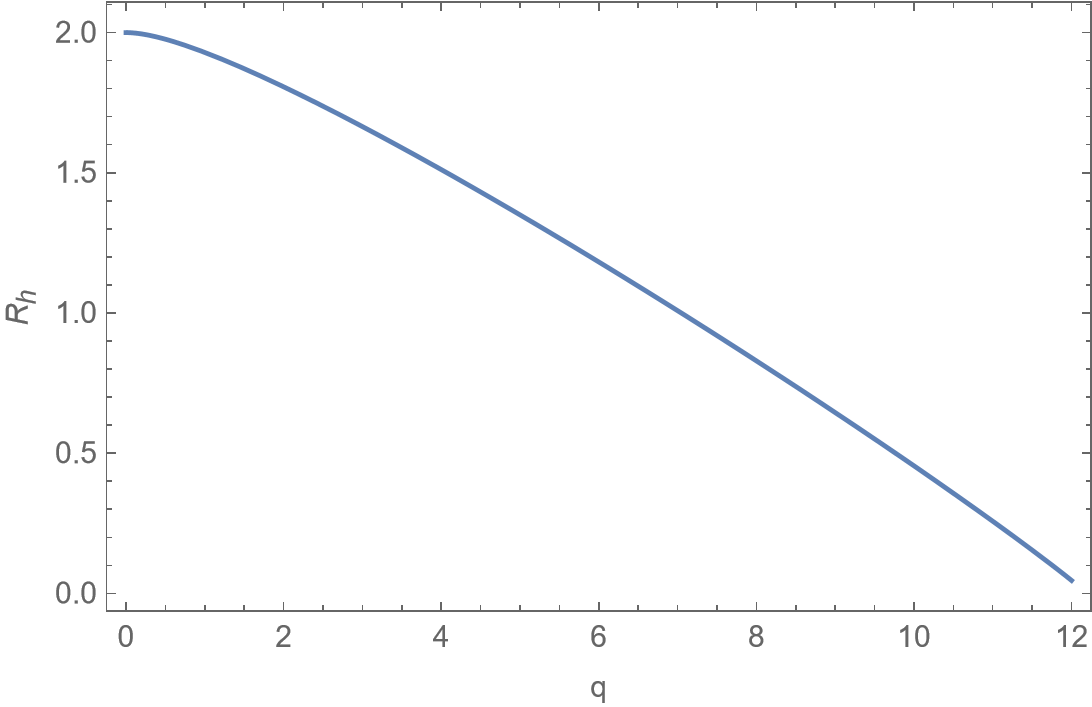} 
\caption{ \it The black hole radius $R_h$ as a function of $q$. Here we set $\gamma =\ft{2}{\sqrt{3}}+1 $, $M=1$.}
\label{Rh-q}
\end{figure}

In the black hole case, the black hole mass $M$ as a function of the physical radius $R$ with different parameters are displayed in Fig.~\ref{rm}. The two plots of the first row are the cases that fixing $\alpha$ and varying $\gamma$. The left one with $\alpha=0.2$ and $\gamma=(\ft{\sqrt{3}}{2}+0.1, \ft{\sqrt{3}}{2}+0.5, \ft{\sqrt{3}}{2}+1)$ correspond to the blue lines from top to down, respectively. The right one with $\alpha=-1$ and $\gamma=(\ft{\sqrt{3}}{2}+0.01, \ft{\sqrt{3}}{2}+0.03, \ft{\sqrt{3}}{2}+0.1)$ correspond to the blue lines from top to down, respectively. The two plots of the second row are the cases that fixing $\gamma$ and varying $\alpha$. The left one with $\gamma=\ft{\sqrt{3}}{2}+0.5$ and $\alpha=(0.3, 0.5, 1.2)$ correspond to the blue lines from top to down, respectively. The right one with $\gamma=\ft{\sqrt{3}}{2}-0.05$ and $\alpha=(-0.3, -0.5, -1.2)$ correspond to the blue lines from top to down, respectively. It is obvious, in a fixed theory, that light black holes have significant deviations from the standard Schwarzschild black hole. To be specific, the hairy black holes possess smaller radii than the same mass Schwarzschild black holes. It follows that the scalar field causes the black holes more compact. A natural question comes to us: how is the smallest hairy black hole we have in this theory?

To address this question, it is convenient take $\alpha=\ft{3(1+q\sqrt{\gamma^2-1})}{2q^3\gamma^2\sqrt{\gamma^2-1}(3\gamma^2-4)}$, i.e. fixing the black hole mass $M=1$. The black hole condition requires $0< q<3(\gamma+\sqrt{\gamma^2-1})$. We display the hairy black hole radii as a function of $q$ in Fig.~\ref{Rh-q}. One can see that the black hole radius tends to $2M$ when $q\to 0$, and tends to $0$ when $q\to3(\gamma+\sqrt{\gamma^2-1})$. It shows that these hairy black holes are always smaller than the standard Schwarzschild black hole and could have infinitesimal areas. The small area, even zero area black holes also arise in string theories \cite{Chowdhury:2024ngg} and other discussions \cite{Holzhey:1991bx,Preskill:1991tb,Huang:2022dfx}.

\section{Energy conditions}

The no-scalar hair theorem states that it is impossible to implement scalar hairs to the black holes in some conditions (see details in a very good review in Ref.\cite{Herdeiro:2015waa}). We obtained it with a specific scalar potential that violates the weak energy condition. 

Interestingly, not all the energy conditions are violated. We examine the null energy condition (NEC) as a warm up. The NEC states that, for any 
null vector $k^\mu$, $k^\mu k^\nu T_{\mu\nu}\geq 0$. With the help of \eqref{EMT}, the contraction gives rise to 
\be
k^\mu k^\nu T_{\mu\nu}=\ft 1 2(k^\mu\partial_{\mu} \phi)^2 \geq0.
\ee
It means the scalar potential will not affect the NEC. \textit{It shows the fact that the normal kinetic term of scalar field is satisfied NEC.} Hence our solution doesn't violate NEC, that is some kind of trivial.

\begin{figure}[t]
\centering
\includegraphics[width=0.4\textwidth]{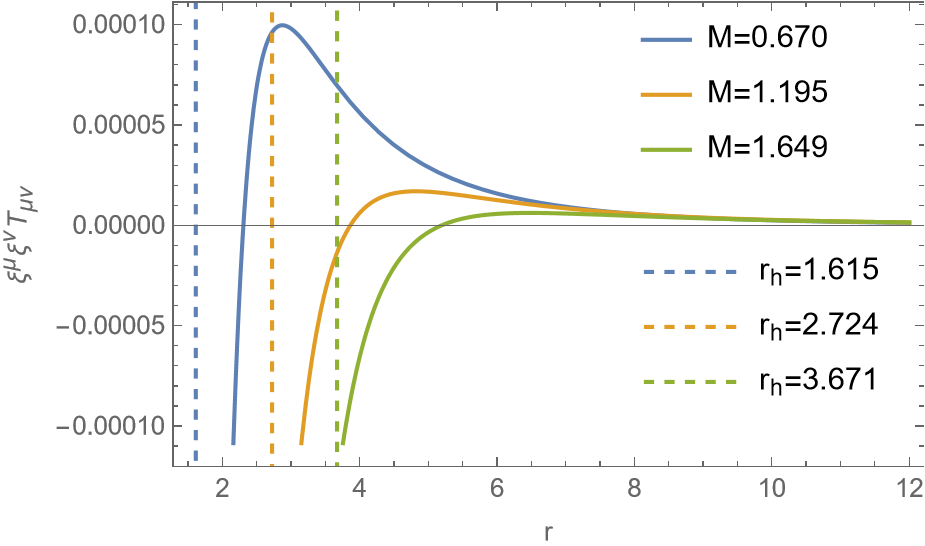} 
\includegraphics[width=0.4\textwidth]{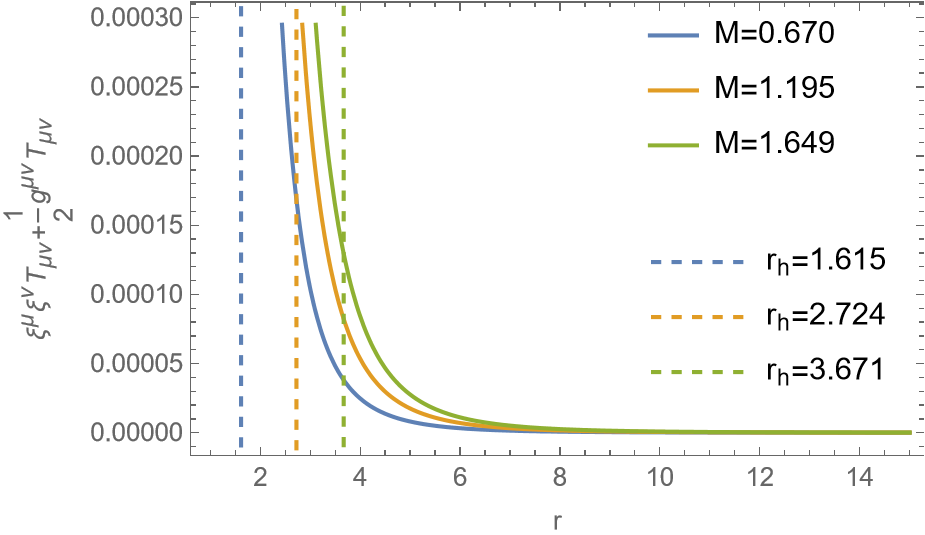} 
\includegraphics[width=0.4\textwidth]{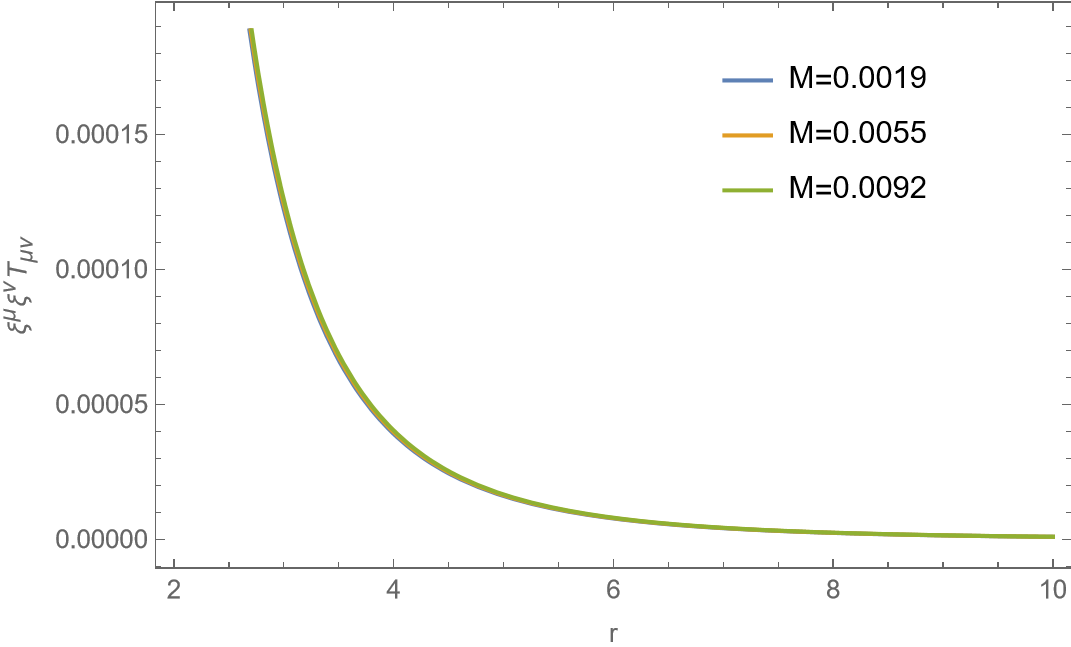} 
\includegraphics[width=0.4\textwidth]{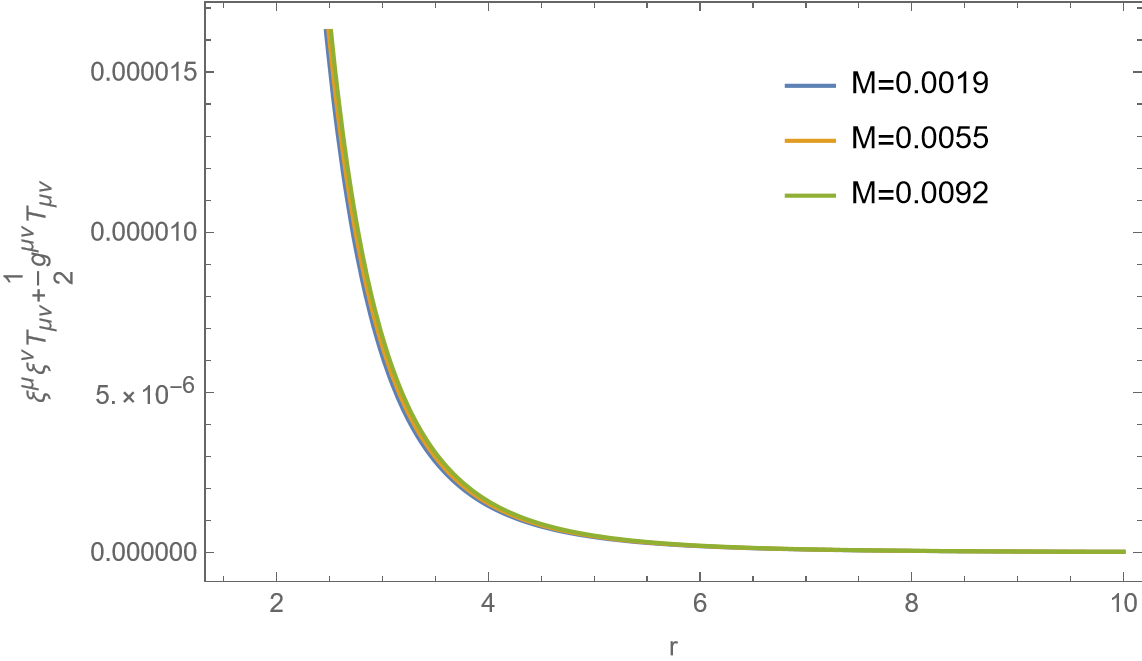} 
\caption{ \it the SEC and WEC condition for the naked singularities and the black hole investigated with $\gamma = 2$, $\alpha=-\beta=4.555$. The first two figures above are black holes, while the last two figures are naked singularities. }
\label{ec}
\end{figure}

As for the weak energy condition (WEC), namely for any time-like vector $\xi^\mu$, it requires $\xi^\mu\xi^\nu T_{\mu\nu}\geq 0$. With the condition $\xi^\mu\xi_\mu=-1$, we have
\be\label{wec}
\xi^\mu\xi^\nu T_{\mu\nu}=\fft 1 2(\xi^\mu\partial_\mu \phi)^2+\fft 1 4(\partial \phi)^2+\fft 1 2 V.
\ee
The inner product of $(\xi^\mu \partial_\mu \phi)^2$ is always positive, so the weak energy condition is determined by the last two terms of \eqref{wec}. We found that WEC is always violated in the vicinity of the black hole horizon. However, the results are always positive for the naked singularity solutions and hence respect the WEC. For instance, we plot the WEC for black holes and naked singularities in Fig.\ref{ec}.

The strong energy condition (SEC) should also be checked, which states that
\be
\xi^\mu\xi^\nu T_{\mu\nu}+\fft{1}{2}g^{\mu\nu}T_{\mu\nu}\geq 0,
\ee
for any time-like vector $\xi^\mu$. In the framework of our theory, the SEC condition equals to 
\be\label{SEC}
\fft 1 2(\xi^\mu\partial_\mu \phi)^2-\fft 1 2 V\geq 0.
\ee
Since $V$ is negative, the SEC is always held. We show an example of SEC in Fig.\ref{ec}.

We may therefore conclude that the naked singularities respect all three energy conditions while the black holes do not.

\section{Quasinormal modes}

Quasinormal modes (QNMs) play an essential role in the black hole perturbation theory. It is a powerful tool to distinguish the gravitational waves emitted by binaries by comparing the QNMs in the ringdown stage. QNMs of black holes are pivotal in unraveling the properties of black holes and gravitational phenomena in the universe. QNMs have complex frequencies, where the real part signifies the oscillation frequency—the inherent rate at which perturbations oscillate around the black hole. Meanwhile, the imaginary part denotes the decay rate, indicating how rapidly the perturbations weaken over time. We would like to investigate the differences between the Schwarzschild black hole and our hairy black hole in this section.

From the theoretical prospects, there are two common perspectives to the black hole perturbations: by imposing an additional test field to the black hole background or perturbing the metric directly\cite{Konoplya:2011qq}. Due to our motivation being mainly to give a qualitative explanation for the time evolution, we take the previous viewpoint and consider only a free scalar as the test field. Furthermore, the fundamental mode is considered in the rest work. 

There are many methods to calculate QNMs, including the WKB approximation method\cite{Konoplya:2003ii,Konoplya:2004ip,Konoplya:2019hlu,Blazquez-Salcedo:2018ipc}, the asymptotic iteration method(AIM)\cite{Jansen:2017oag}, the spectral method\cite{Blazquez-Salcedo:2023hwg,Khoo:2024yeh,Chung:2023zdq,Chung:2023wkd,Azad:2024axu}, the finite difference method \cite{Ou:2021efv,Huang:2021qwe,Guo:2021enm,Guo:2022umh}, and so on. We adopt the time-evolution method by numerically solving the E.O.Ms and use the Pronny method to extract the QNM's frequencies.

\subsection{Test field and the effective potential}

We consider a massless scalar as the test field, in the black hole background. The E.O.M of the scalar field is
\be\label{eq1}
\Box\psi =\fft{1}{\sqrt{-g}}\partial_\mu\bigg(\sqrt{-g}g^{\mu\nu}\partial_\nu\psi(t,r,\theta,\varphi)\bigg)=0.
\ee
As a static and spherically space-time, we can separate the scalar field $\psi$ into spherical harmonics $Y_{lm}$ and radial part $\Phi(t,r)$, namely
\be\
\psi(t, r,\theta, \varphi)=\sum_{l, m}\fft {\Phi(t, r)}{R(r)} Y_{lm}(\theta, \varphi).
\ee
Thus, the above equation \eqref{eq1} reduces to
\be\label{eq2}
-\fft{\partial^2\Phi(t, r)}{\partial t^2}+hf\fft {\partial^2\Phi(t, r)}{\partial r^2}+\fft{1}{2}(f h'+h f')\fft{\partial\Phi(t, r)}{\partial r}-V(r)\Phi(t, r)=0,
\ee
with the effective potential
\be\label{vqnm}
V(r)=\fft {l (l+1)}{R^2}h+\fft{1}{2R}R'(hf)'+\fft{h f R''}{R}. 
\ee
The tortoise coordinate $r_{*}$ is employed by the transformation
\be
dr_{*}=\fft{dr}{\sqrt{hf}}.
\ee
With the help of tortoise coordinate, one can map the radial region from $r\in (r_h, +\infty)$ to $r_* \in (-\infty, +\infty)$. The scalar field equation \eqref{eq2} reduce to a Schr\"odinger-like equation
\be\label{sch}
-\fft{\partial^2\Phi(t,r)}{\partial t^2}+\fft {\partial^2\Phi(t,r)}{\partial r_{*}^2}-V(r)\Phi(t,r)=0.
\ee

\begin{figure}[h]
\centering
\includegraphics[width=0.5\textwidth]{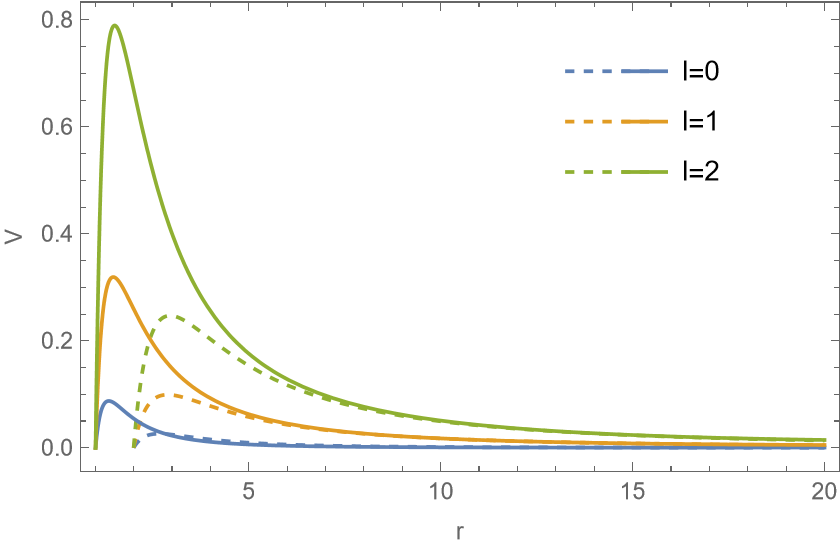}
\caption{ \it 
Effective potentials for various values of angular momentum. The dashed line corresponds to Schwarzschild black hole and the solid line to the hairy black hole. We set $\gamma =\ft{2}{\sqrt{3}}+1 $, $M=1$, $q=7$.}
\label{plotv}
\end{figure}

Before we perform the numerical algorithm to solve the equation \eqref{sch}, it is important to analyze the behaviors of the effective potential \eqref{vqnm} of the hairy black holes. It always exhibits a single peak, similar to other typical black holes. We present some examples of the effective potential in Fig.\ref{plotv}. The effective potential is zero at the event horizon, then ascends to reach a peak, and subsequently diminishes, indicating distinct behavior in quasinormal modes (QNMs). And, the peak of the hairy black hole's potential increases with the angular number $l$, while shifting to the right, resembling the characteristics of a Schwarzschild black hole. Furthermore, the dependence of the hairy black hole on the angular number $l$ is similar to that of the Schwarzschild black hole.

Nonetheless, one of the distinctions between this hairy black hole and the Schwarzschild black hole is the high of peaks in the effective potential for the hairy black hole is greater than that of the Schwarzschild black hole for the same angular momentum parameter. This feature affects the oscillated frequencies and the damping times in their QNMs.

\subsection{Time domain integration}

Now we used the Finite difference method to solve the Schr\"odinger-like wave equation. To begin with, we discretize the coordinates $t=i\Delta t$ and $r_{*}=i\Delta r_{*}$, then the equation \eqref{sch} becomes
\be
-\fft {\Phi(i+1,j)-2\Phi(i,j)+\Phi(i-1,j)}{\Delta t^2} +\fft {\Phi(i,j+1)-2\Phi(i,j)+\Phi(i,j-1)}{\Delta r_{*}^2}-V(j) \Phi(i,j)=0.
\ee
The initial condition that we used is the Gaussian distribution \cite{Liu:2020qia}
\begin{eqnarray}
&&
\begin{cases}
&\Phi(t=0, r_{*}) = e^{-\fft{(r_{*}-a)^2}{2 b^2}},\\
    &\Phi(t<0, r_{*})=0,
\end{cases}
\end{eqnarray}
choosing $b=1$ and determining the value of a accordingly in this work. Consequently, the equation of $\Phi$ reduces to
\be
\Phi(i+1,j)=-\Phi(i-1,j)+\bigg(2-2 \fft {\Delta t^2}{\Delta r_{*}^2}-\Delta t^2 V(j)\bigg)\Phi(i,j)+ \fft {\Delta t^2}{\Delta r_{*}^2}\bigg(\Phi(i,j+1)+\Phi(i,j-1)\bigg)\,.
\ee
Inherently, solving the above equation requires appropriate boundary conditions. In black hole background, there are only ingoing waves at the horizon and outgoing waves at the infinity. It implies that the boundary conditions are
\be 
\partial_{t}\Phi = \pm \partial_{r_{*}} \Phi,   \qquad \text{for}\quad r_{*}\to \pm \infty.
\ee
We take $\ft{\Delta t}{\Delta r_{*}}=0.5$ to  satisfy the von Neumann stability condition, which requires $\ft{\Delta t}{\Delta r_{*}}<1$.

\begin{figure}[h]
\centering
\includegraphics[width=0.32\textwidth]{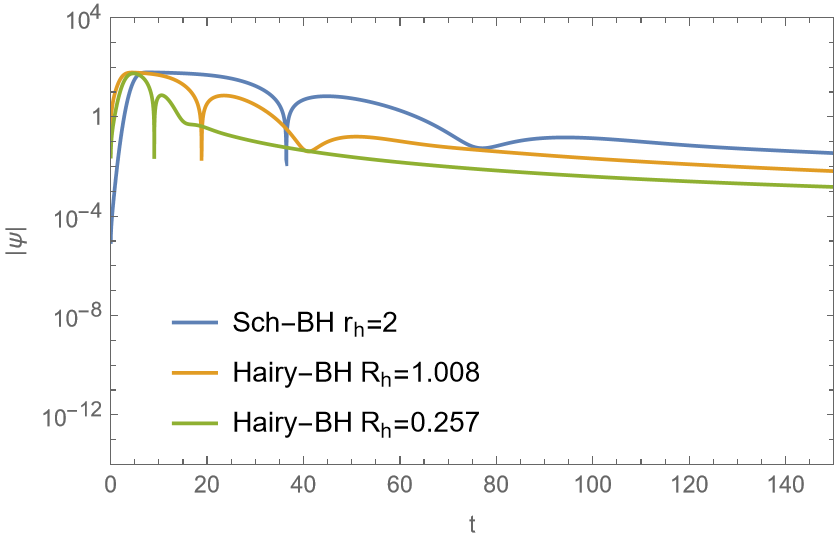} 
\includegraphics[width=0.32\textwidth]{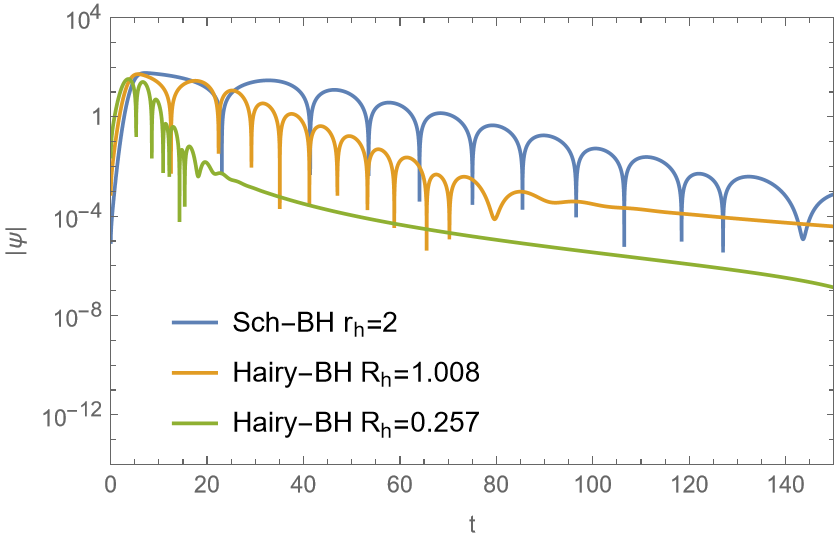}
\includegraphics[width=0.32\textwidth]{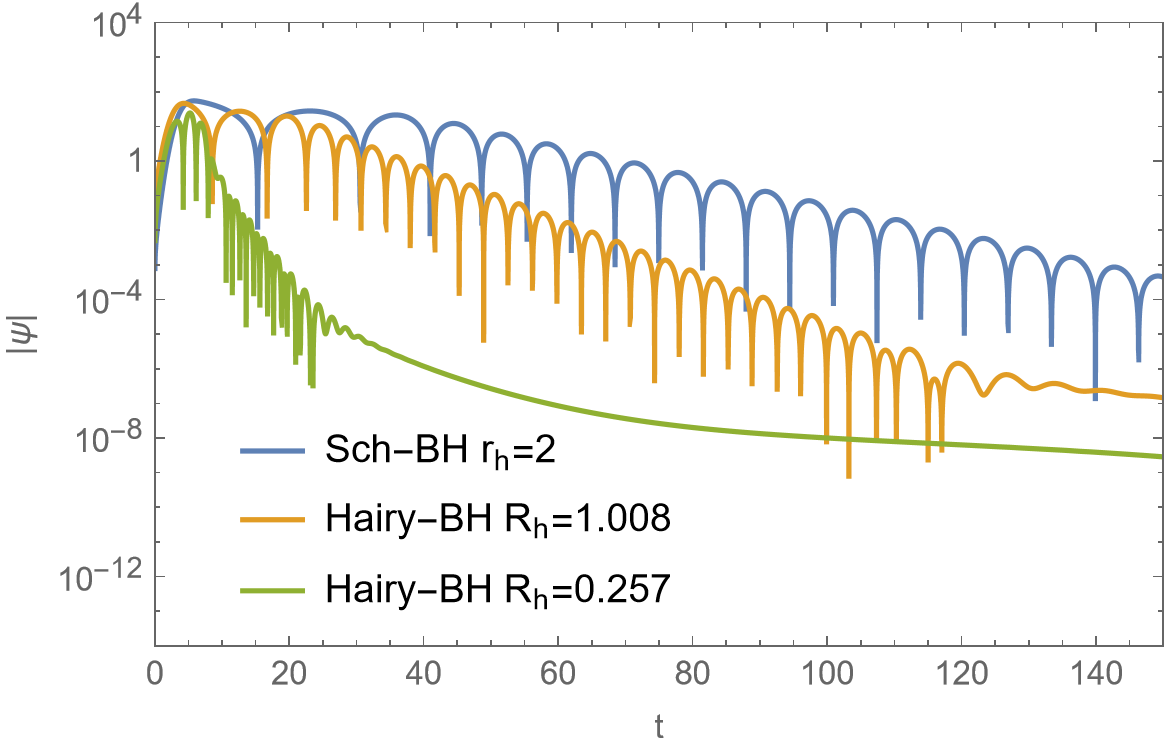}
\caption{ \it Time-domain plots for the Schwarzschild black hole and the hairy black hole. The left plot corresponds to $l=0$, the middle one to $l=1$, and the right one to $l=2$. Here, we take $\gamma =\ft{2}{\sqrt{3}}+1 $, $M=1$.}
\label{plotphi}
\end{figure}

\begin{figure}[h]
\centering
\includegraphics[width=0.4\textwidth]{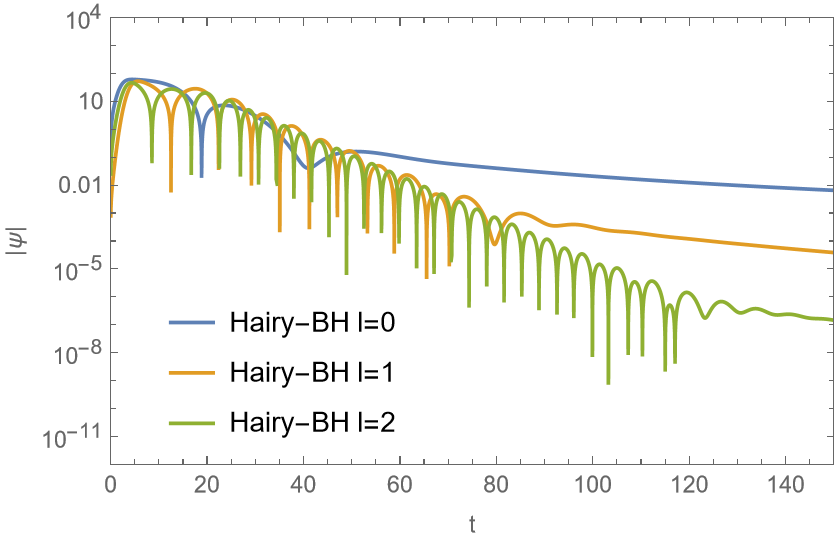} 
\includegraphics[width=0.4\textwidth]{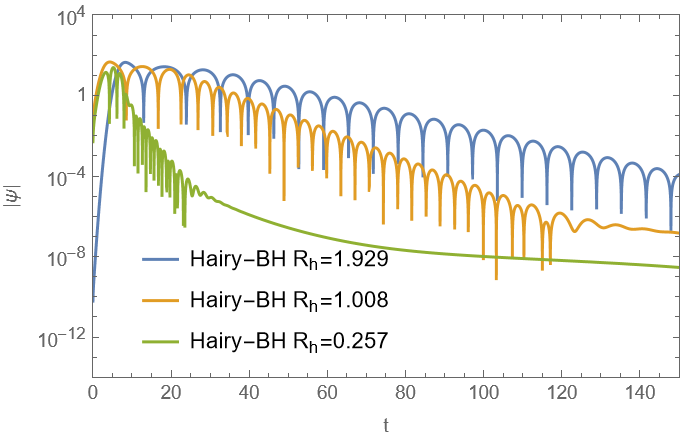}
\caption{ \it Time-domain plots for the hairy black hole. The left plot corresponds to $R_{h}=1.008$, and the right plot corresponds to $l=2$. Here, we take $\gamma =\ft{2}{\sqrt{3}}+1 $, $M=1$.}
\label{plotphi2}
\end{figure}

We present the time-domain plots for the Schwarzschild black hole and the hairy black hole in Fig.\ref{plotphi} and Fig.\ref{plotphi2}. The time-domain plots tell us the black holes remain stable under the influence of a massless scalar field. Both the hairy black hole and the Schwarzschild black hole exhibit a decelerated decay rate as the angular number $l$ increases. Furthermore, when considering a fixed angular number $l$, the hairy black hole with a smaller black hole radius displays a faster decay rate compared to the Schwarzschild black hole in Fig.\ref{plotphi}. This suggests that the black hole radius plays a significant role in the long-term behavior and characteristics of these black holes.

Fig.\ref{plotphi2} presents the variation of the frequencies of hairy black holes with respect to the angular number under the same set of parameters. Moreover, it shows the variation of hairy black holes with respect to the black hole radius for $l=2$. The time-domain plot for $l=2$ is also included as it plays a dominant role in the ringdown signals of gravitational waves. We found the decay is observed to decelerate further with an increase in both $l$ and $R_{h}$.

\subsection{QNM's frequencies}

It is a standard method to use the Pronny Method to extract the QNM's frequencies. We compared the QNM's frequencies of our obtained Schwarzschild black holes with those from previous studies in Ref.\cite{Konoplya:2023moy} and found that we were able to achieve precision up to the second decimal place. This demonstrates the effectiveness of our computational method. We focused primarily on the variations of the real and imaginary parts of the frequency with respect to the horizon $R_{h}$ for the hairy black hole, as well as the variations of these parts with respect to horizon $R_{h}$ for different angular number $l$.

In detail, the new hairy black hole with $l=1,2$ are shown in Table.\ref{hairytable}. A larger real part implies a higher oscillation frequency, while a larger imaginary part indicates faster decay and a quicker disappearance of perturbations. The data in the table reveals a notable trend: as the black hole radius $R_{h}$ increases, the real part of its frequency decreases, indicating that the larger black hole radius is more effective in suppressing oscillations. Furthermore, the imaginary part of the frequency also decreases with an increase of black hole radius, meaning decaying more slowly and providing evidence for the black hole's stability under external perturbations. Besides, the standard Schwarzschild black hole radius is always bigger than the hairy black hole with the same mass, so hairy black holes have higher oscillation frequency and decay faster. These findings align with the conclusions drawn from the previous time-domain analysis.

\begin{table}[htbp]
    \centering
    \begin{tabular}{|c|c|c|c|}
        \hline
        $R_{h}$ & $l=1$  & $l=2$ & $l=3$ \\
        \hline
        0.512 &  0.980555 -0.329293i & 1.615452 -0.325563i  &  2.248509 -0.320262i \\
  1.018 & 0.526765 -0.173979i  & 0.862807 -0.172580i &  1.206043 -0.172277i \\
   1.511 &0.366889 -0.121971i& 0.601491 -0.120219i &  0.841138 -0.120160i \\
   2$(Sch)$ & 0.293421 -0.097667i& 0.483155 -0.096735i &  0.675133 -0.096580i \\
        \hline
    \end{tabular}
    \vspace{0.5em}
    \caption{The quasinormal modes of the scalar field in the hairy black hole and Schwarzschild black hole. Here we set $\gamma =\ft{2}{\sqrt{3}}+1 $, $M=1$.}
    \label{hairytable}
\end{table}

\section{Conclusions}

The black hole no-hair theorem has ruled out black holes with scalar hair in Einstein's gravity with certain conditions. However, one can bypass the theorem to find out more black hole solutions by diving into modified gravities or by ignoring the conditions of the theorem. We obtained such hairy black hole solutions in the latter scheme, namely got solutions in the Einstein-scalar theory. We found these solutions can be asymptotic to Minkowski, (A)dS spacetimes, depending on the effective cosmological constant $\Lambda$. 

To the interests of astrophysics, we work on the details of the asymptotic Minkowski solutions only. As we showed in the paper, the solution can depict single-horizon black holes or time-like naked singularities. This implies the solution is not a simple generalization of Schwarzschild metric. In fact, it can’t reduce Schwarzschild metric smoothly by taking $q\to 0$. Furthermore, we found the hairy black holes can possess arbitrary small areas, and their areas are always smaller than the same mass Schwarzschild black holes.

We examined the energy conditions of the hairy black holes and found they violate only the weak energy condition but respect the null and strong energy conditions. In contrast, the time-like naked singularities respect all three energy conditions.

As a hairy black hole with a single horizon, it is an interesting direction to do the comparison with Schwarzschild black hole. We considered a test scalar field in the black hole background and evaluated its time evolutions, which gives rise to the QNMs. We compared the QNM’s frequencies between the hairy black hole and Schwarzschild black hole with the same mass. The results suggest that, even the tendencies are quantitively similar, one can still distinguish the hairy black holes by its QNMs. In fact, the hairy black holes have smaller areas compared with the same mass Schwarzschild black hole, which leads to higher oscillation frequencies and shorter decay times.

We did not investigate the asymptotic (A)dS hairy black holes in this work. Based on the importance of the non-asymptotic flat black holes in theoretical studies, it deserves further studies. For future works, one can also generalize the solution to the rotation one. It has the potential to show more differences between hairy black holes with normal black holes. Our solution also provides an analytical example to the applications of black hole physics.

\section*{Acknowledgment}
XPR and HH gratefully acknowledge support by the National Natural Science Foundation of China (NSFC) Grant No.~12205123 and Jiangxi Provincial Natural Science Foundation
with Grant No.~20232BAB211029 and by the Sino-German (CSC-DAAD) Postdoc Scholarship Program, No.~2021 (57575640).
This work is also partially supported by the National Natural Science Foundation of China (No. 11873025).

\section*{Appendix A}\label{app}

In this appendix, we present how to obtain the theory by a disformal transformation of an Einstein-phantom scalar theory in Ref.\cite{Huang:2020qmn}.

A trick we used is noting that the following transformation of metric
\be
 g_{\mu\nu}\to K g_{\mu\nu},
\ee
with $K$ is constant, leads to
\bea
\sqrt{-g}&\to&K^2\sqrt{-g}\,,\nn\\
(\partial \phi)^2&\to&K^{-1}(\partial \phi)^2\,,\nn\\
R&\to& R\,,\nn\\
V&\to&V.
\eea
Then the Lagrangian \eqref{lag} becomes
\be
{\cal L}=K^2 {\cal L},
\ee
which do not change the E.O.Ms. For our purpose, we take 
\be
K=e^{-\sqrt{\gamma^2-1}\fft{i\pi}{\gamma}}.
\ee
Together with the identity $\arctan x=i\text{arctanh} \ft{i}{x}+\ft{\pi}{2}$, we therefor obtain the new scalar potential and normal scalar 
\bea
V&=&e^{-\sqrt{\gamma ^2-1} \phi }\alpha\bigg((2-3\gamma^2)\text{cosh}(\gamma\phi)-3\gamma\sqrt{\gamma^2-1}\text{sinh}(\gamma\phi)-2\bigg)\nn\\
&&-e^{\sqrt{\gamma^2-1} \phi }\beta\bigg(-(2-3\gamma^2)\text{cosh}(\gamma\phi)-3\gamma\sqrt{\gamma^2-1}\text{sinh}(\gamma \phi)+2\bigg),\nn\\
\phi &=&\fft{2}{\gamma}\text{arctanh}(\fft{q\gamma}{\sqrt{r^2+q^2\gamma^2}}).
\eea
By applying the corresponding transformation to the metric function, one can obtain the black hole solution that we presented in \eqref{solution}.

\end{document}